\documentclass{article}
\usepackage{spconf,amsmath,graphicx}
\usepackage{color}
\usepackage{booktabs}
\usepackage{threeparttable}

\title{coronary wall segmentation in CCTA scans via a hybrid net with contours regularization}

\name{Author(s) Name(s)\thanks{Thanks to XYZ agency for funding.}}
\address{Author Affiliation(s)}

\name{ 
\begin{tabular}{@{}c@{}}
Kaikai Huang$^{\star \dagger}$ \qquad 
Antonio Tejero-de-Pablos$^{\star \dagger}$ \qquad 
Hiroaki Yamane$^{\dagger \star}$ \qquad 
Yusuke Kurose$^{\star \dagger}$ \\
\qquad Junichi Iho$^{\ddagger}$ \qquad 
Youji Tokunaga$^{\ddagger}$ \qquad 
Makoto Horie$^{\ddagger}$ \qquad 
Keisuke Nishizawa$^{\ddagger}$  \\
\qquad Yusaku Hayashi$^{\ddagger}$ 
\qquad Yasushi Koyama $^{\ddagger  \dagger}$ 
\qquad Tatsuya Harada$^{\star \dagger}$
\end{tabular}}

\address{$^{\star}$ The University of Tokyo, Bunkyo 113-8654, Tokyo, Japan \\
    $^{\dagger}$ RIKEN Center for Advanced Intelligence Project, Chuo 103-0027, Tokyo, Japan \\
    $^{\ddag}$ Sakurabashi Watanabe Hospital, Kita 530-0001, Osaka, Japan}
    
\begin{document}
%
\maketitle

\begin{abstract} \label{sec:abstract}
Providing closed and well-connected boundaries of coronary artery is essential to assist cardiologists in the diagnosis of coronary artery disease (CAD). Recently, several deep learning-based methods have been proposed for boundary detection and segmentation in a medical image. However, when applied to coronary wall detection, they tend to produce disconnected and inaccurate boundaries. In this paper, we propose a novel boundary detection method for coronary arteries that focuses on the continuity and connectivity of the boundaries. In order to model the spatial continuity of consecutive images, our hybrid architecture takes a volume (i.e., a segment of the coronary artery) as input and detects the boundary of the target slice (i.e., the central slice of the segment). Then, to ensure closed boundaries, we propose a contour-constrained weighted Hausdorff distance loss. We evaluate our method on a dataset of 34 patients of coronary CT angiography scans with curved planar reconstruction (CCTA-CPR) of the arteries (i.e., cross-sections). Experiment results show that our method can produce smooth closed boundaries outperforming the state-of-the-art accuracy.
\end{abstract}


\section{Introduction} \label{sec:intro}

Coronary artery disease (CAD), the narrowing or blockage of the coronary arteries, is one of the leading causes of death around the world. It is usually caused by atherosclerosis (i.e., the build up of plaques on the inner walls of arteries) and can restrict blood flow to the heart muscle by physically clogging the artery. This is the cause of abnormal artery functions and other related diseases such as angina, heart attack, and ischemia. The CAD diagnosis is a complicated and time-consuming procedure requiring expertise of well-trained cardiologists. In particular, precise blood vessel wall segmentation of the coronary artery plays a fundamental role in the whole procedure. With the advances of automatic vessel extraction techniques, centerlines of the main coronary arteries can be extracted given a volume of CCTA scans of the heart. Then a curved planar reconstruction (CPR) is operated to extract cross-section slices along the curved centerline and rearrange them into a new volume (i.e., CPR artery), from which cardiologists will segment the coronary wall.

This coronary wall segmentation task can be solved with a boundary detection approach by detecting both the inner boundary (the boundary between vessel lumen space and extra-vascular space) and the outer boundary (the boundary between extra-vascular space and the background) of the given CPR artery. Most typical edge detection methods (e.g., Canny edge detector, Robertfilter, Sobel filter and Snakes~\cite{marquez2014amorph}) use gradient-based features for boundary identification, which is also applicable to coronary artery boundary detection. Besides for that, Mazhar B. Tayel applied a modified fast algorithm using the Lucy-Richardson method and the 2D median filter to determine the inner and outer contours of given coronary artery~\cite{tayel2017449, tayel2014Anas}. Merkow used a forest of decision trees to classify each voxel into the outer boundary and non-outer boundary~\cite{merkow2015strctual}. However, these methods are sensitive to noise variance and the fine-tuning of hyper-parameters is complex. In many cases, manual correction of the result is required to obtain the exact object area, therefore increasing the workload of cardiologists. 

The most recent works use convolutional neural netwroks (CNNs) and other deep-learning techniques for boundary detection in a medical image, among those the most commonly used methods are U-Net~\cite{ronneberger2015unet} and its extensions~\cite{Zhang2018RoadEB, oktay2018attentionunet, Alom2018RecurrentRC, kohl2018probunet}. The state-of-the-art also proposes leveraging the consistency between boundary detection and segmentation results. A boundary-enhanced segmentation network (BESNet)~\cite{odahirohisa2018besnet} employs two decoding paths, for boundary detection and segmentation, respectively. A boundary regularized deep convolutional encoder-decoder network (ConvEDNet)~\cite{bian2017convednet} was proposed, featuring a deep boundary supervision strategy to regularize the feature learning for better robustness to speckle noise and shadowing effect. However, most of those methods failed to consider the spatial dependency between target image (i.e., the image for boundary detection) and the preceding and posterior images. In order to leverage spatial dependency, a 3D volume-to-volume CNN architecture I2I-3D~\cite{merkow2016i2i3d} was used to predict boundary location in the given vascular volume. However, network architectures with fully-3D up-convolutions in the expansion path lead to high computational complexity, thus limit its actual application. Moreover, none of the aforementioned methods deals with ensuring connected closed boundaries, a problem that is particularly challenging in coronary artery wall detection.

In this paper, we propose a novel method for automatic coronary wall segmentation by detecting boundaries (i.e., the inner and outer boundaries) of the coronary artery and achieve the state-of-the-art accuracy. The contributions of this work are two-fold:
\begin{itemize}
\item[.] We propose a hybrid network architecture comprising 3D encoder and 2D decoder. This allows leveraging the spatial continuity of consecutive slices to detect the boundary of the target slice. Compared with fully-3D approaches, our architecture alleviates the computational cost, while preserving the integrity of the output.
\item[.] Based on the weighted Hausdorff distance (WHD) loss~\cite{wdhfjavier2018}, we propose a novel contour-regularized WHD loss, which ensures connected closed boundaries by applying a contour regularization to the network output.

\end{itemize}

\section{coronary wall segmentation with contour regularization} \label{sec:proposal}

\begin{figure*}[!tb]
\centering
\includegraphics[width=\textwidth]{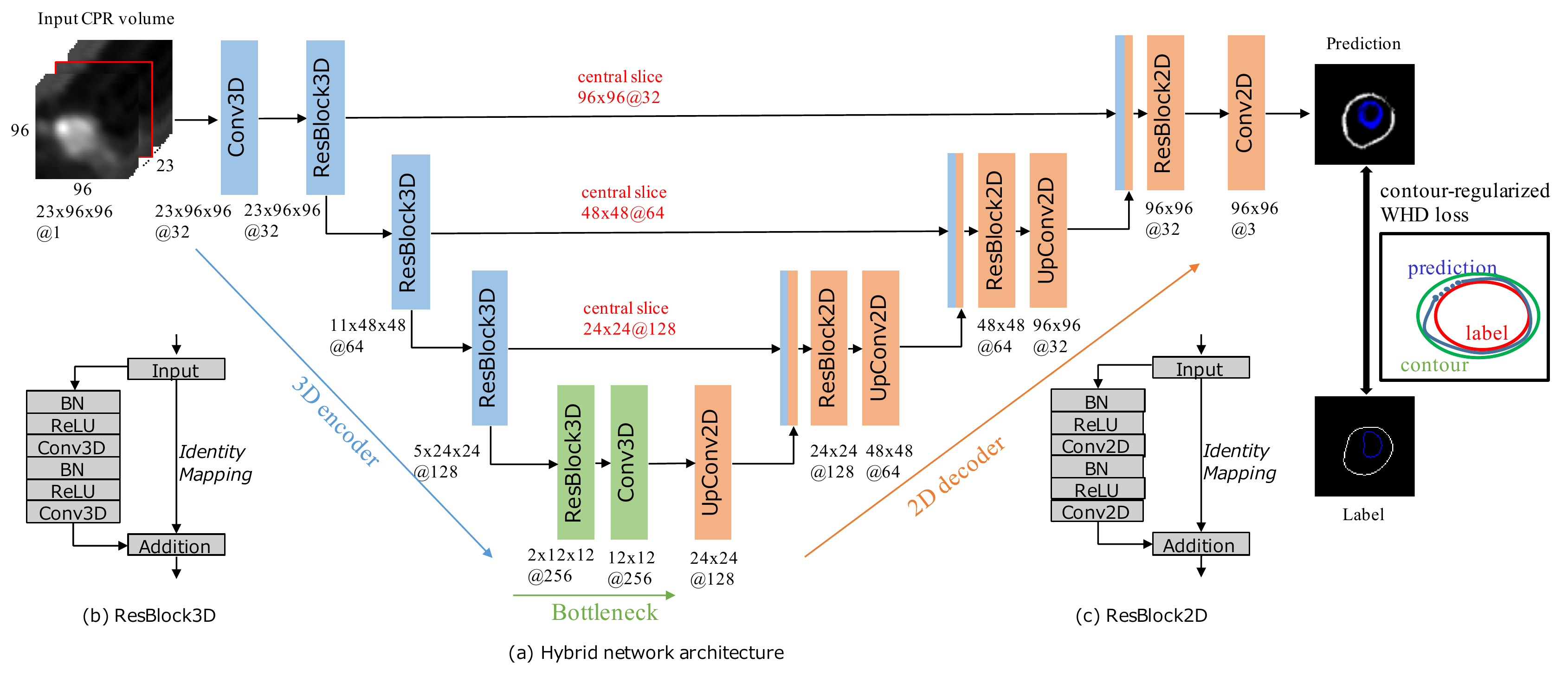}
\caption{Overview of our method. (a) The hybrid network comprises the encoder path with 3D residual blocks, the bottleneck layers, and the decoder path with 2D residual blocks and 2D up-convolutions. (b,c) Architecture of the 3D residual block and the 2D residual block, respectively} \label{fig: hru}
\end{figure*}

Fig.~\ref{fig: hru} shows an overview of the proposed method, which classifies each pixel of the target image into the inner boundary, the outer boundary or background. Given an input CPR volume with central slice as the target image, the hybrid network outputs the probability map (i.e., probabilities of being the inner boundary, the outer boundary or background for each pixel) of the target image. Then, contour-regularized WHD loss between the output and the ground truth (GT) boundaries is calculated and minimized through back propagation. This procedure is repeated over all training samples until convergence, and the final prediction for each pixel corresponds to the class with the highest probability over all three channels.

\subsection{Hybrid Network Architecture \label{sec: hybridresunet}}

In order to detect boundaries from an input image $I_t$ (i.e., the $t$-th cross-section slice of the whole artery $I$), we use a volume $V=\{I_{t-N/2}\ldots I_t\ldots I_{t+N/2}\}$ centered at $I_t$ with $N+1$ consecutive slices (e.g., 23 in our case) as input. For the special case that $I_t$ is in the first $N/2$ slices of the artery (i.e., $t\in[1,N/2]$), we make it the center by padding a reflection of the input. The same reflection padding is used when $I_t$ approaches at the final $N/2$ slices of the artery.

As shown in Fig.~\ref{fig: hru}(a), the proposed hybrid network architecture is composed of a 3D encoder, the bottleneck layers, and a 2D decoder. For a given input volume $V$, the encoder processes it with one 3D convolution, followed by 3 consecutive 3D residual blocks, comprising two blocks of batch normalization, ReLU activation and 3D convolution as shown in Fig.~\ref{fig: hru}(b). Then, the bottleneck layers further processes the encoder output with one 3D residual block and one 3D convolution, under which the number of slices in the feature map will be reduced to one. Finally, symmetrical to the encoder, the decoder comprises three groups of 2D up-convolution and 2D residual block (as shown in Fig.~\ref{fig: hru}(c)) followed by one 2D convolution. Moreover, inspired by the residual U-Net~\cite{Zhang2018RoadEB}, we also apply skip connections for direct feature propagation. Considering that the 3D feature maps in the encoder cannot be directly concatenated to the 2D feature maps in the decoder due to size mismatch, we only propagate the central slice of the feature maps in the encoder to the corresponding feature maps in the decoder. Our proposed hybrid structure mainly has two advantages. First, the preceding and posterior slices of the target slice are leveraged as a reference to improve boundary detection accuracy. Second, the overall computational cost of our 3D-2D architecture is much lower than that of a fully-3D one.

\subsection{Contour-regularized Weighted Hausdorff Distance Loss} \label{sec:WHD_contour_reg}
The average Hausdorff distance (AHD) is a representative metric to evaluate boundary difference between the ground truth boundary points and the predicted boundary points. Based on AHD, the weighted Hausdorff distance (WHD)~\cite{wdhfjavier2018} loss was proposed as the loss function to train a neural network:
\begin{equation} \label{equ: whd}
\begin{split}
d_{H}(p, Y) & = \frac{1}{|\hat{X}| + \epsilon} \sum_{x \in \Omega} p_x \min_{y \in Y} d(x, y) \\
    & + \frac{1}{|Y|} \sum_{y \in Y} \min_{x \in \Omega} \frac{d(x, y) + \epsilon}{p_x^{\alpha} + \frac{\epsilon}{d_{max}}},
\end{split}
\end{equation}
where $|\hat{X}| = \sum_{x \in \Omega} p_x$. $\Omega$ is the pixel coordinate space, $x$ any pixel point in $\Omega$ and $p_x \in [0, 1]$ the probability of being a boundary point at $x$. $Y$ is the GT pixel coordinate space and $|Y|$ the number of pixels in $Y$. $d(x, y)$ is the AHD between point $x$ and point $y$, and $d_{max}$ the allowable maximum distance between any two points in $X$ and $Y$. $\epsilon$ is the infinity minimal term for numerical stability during calculation, and $\alpha$ is the order of penalty for $p_x$ when the actual boundary points are miss-estimated. In our experiment, we set $\epsilon = 10^{-6}$ and $\alpha = 4$. 

A serious problem of WHD loss is that it does not ensure obtaining a closed boundary with one-pixel thickness. Therefore, we propose restricting the network output within the space between the GT boundary and a closed contour regularization (as shown in Fig.~\ref{fig: hru}), where the closed contour is a post-processing of the network output. Therefore, the final loss (i.e., contour-regularized WHD loss) is the sum of $d_{H}(p, Y)$ (i.e., the WHD loss between network output $p$ and the GT boundary $Y$) and $d_{H}(p, S)$ (i.e., the WHD loss between $p$ and the contour regularization $S$):
\begin{equation} \label{equ: snakewhd}
d(p, Y, S) = d_{H}(p, Y) + d_{H}(p, S),
\end{equation} 
Considering that the outputs are not stable in the beginning of the training, the contour regularization is not applied to the first ten epochs. For the following 40 epochs, the contour-regularized loss is updated under the rules below until the global minimum is achieved.
\begin{itemize}
    \item [1] For a given probability map $p$, calculate $d_{H}(p, Y)$.
    \item [2] Convert $p$ to a closed contour $S$ via border following algorithm~\cite{suzuki85findcontour} or morphological Snakes~\cite{marquez2014amorph} if a closed contour is not obtained, and calculate $d_{H}(p, S)$.
    \item [3] Calculate $d(p, Y, S)$ and update the network parameters
\end{itemize}

With enough iterations, the outline of the predicted probability map $p$ and $S$ will be close enough to each other, thus we can use $S$ as the final prediction. Besides, $S$ is a closed boundary with one-pixel thickness.

For each cross-section slice, there exist both the inner boundary and the outer boundary, where the outer boundary is more challenging to detect due to partial volume effect and relatively low HU contrast near extra-vascular and background interface. Due to this difference, we calculate the WHD loss w.r.t. the inner and outer boundaries separately and sum them up under different ratios to obtain the final loss $d(p, Y, S) = r_i d(p_i, Y_i, S_i) + r_o d(p_o, Y_o, S_o)$,
where $i$ is for inner, $o$ for outer, and $r_i + r_o = 1$. We chose the optimal ratio (i.e., 1:5) and the number of slices in the input volume (i.e., 23) through extensive experiments.

\section{Experimental Results} \label{sec: results}

\begin{figure*}[tb]
\centering
\includegraphics[width=\textwidth]{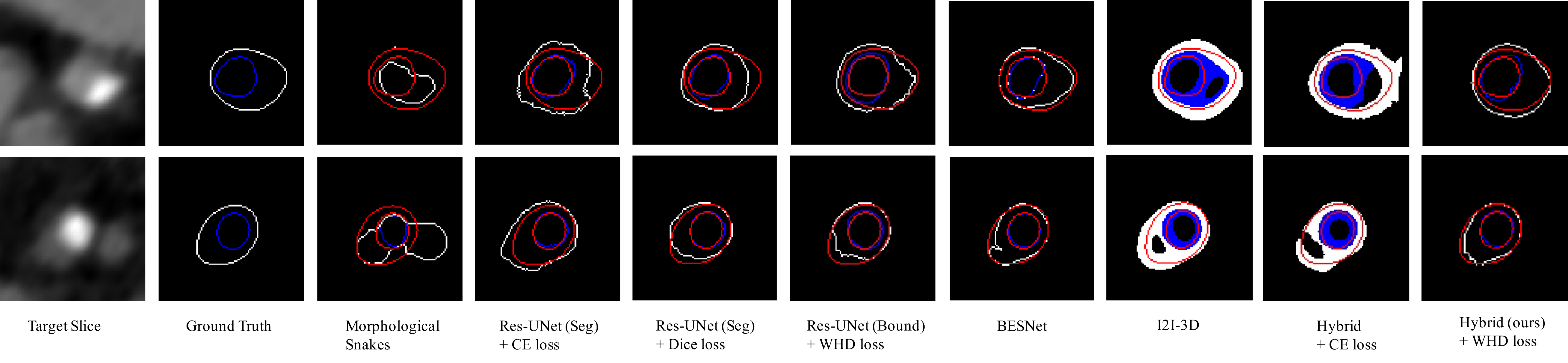}
\caption{Qualitative results comparison between our method and the others. Ground truth boundaries are drawn in red color, whereas the detected inner boundary is marked in blue and the outer boundary in white.} \label{fig: results_comp}
\end{figure*}

We evaluated our method on a non-public CCTA-CPR dataset with three artery volumes for each patient: right coronary artery (RCA), left anterior descending (LAD) coronary artery and left circumflex (LCX) coronary artery. The CCTA scans were taken via a 256-slice MDCT scanner (Brilliance iCT, Philips Medical Systems) with 128 mm $\times$ 0.625 mm collimation, 270 ms gantry rotation time, and 320-840 mA effective tube current (higher values in obese patients). Cardiac images from the split of the pulmonary artery to the apex of the heart were acquired at the 75\% cardiac phase (45\% for heart rate higher than 75) of one RR interval on ECG. Images were reconstructed to an in-plane resolution ranging from 0.38 to 0.56 mm, with 0.9 mm slice thickness and 0.45 mm slice increment. Finally, comprehensive Cardiac Analysis software was used to stretch multi-planner reformation of coronary arteries into CPR views. Coordinates that did not match the original image (e.g., proximal coordinates including the aorta and distal coordinates of unreliable small and/or bend vessels on inadequate contrast enhancement) were excluded for usage, summing a total of 33,676 effective instances from 34 patients. Among them, 25 patients (24,727 slices) were randomly selected for training, 4 for validation (3,231 slices) and the remaining 5 (5,718 slices) for test. All slices were processed with the same pre-processing and augmentation: Hounsfield Units (HU) mapped to a floating point value between 0 and 255, random central crops (with size ranging from 192 to 256) to remove irrelevant information while remaining all artery voxels, image resize (to $96\times96$) to alleviate computational burden, random rotations (with every 90 degree), random horizontal \& vertical flip. Adam optimizer with weight decay 0.0005 and initial learning rate 0.001, decayed by 0.9 for every 10 epochs is used to train the hybrid network. For both our proposal and related works, the training procedure is iterated for 50 epochs. 

We calculated the Dice similarity coefficient (DSC), Average boundary distance (ABD), and the 95 percentile Hausdorff distance (HD95) metrics, respectively for comparison with the related works. We employed several representative methods for comparison, including the morphological Snakes~\cite{marquez2014amorph}, Res-UNet~\cite{Zhang2018RoadEB} with Dice and Cross Entropy loss respectively for segmentation (boundaries are the outlines of the segmented coronary wall), Res-UNet~\cite{Zhang2018RoadEB} with WHD loss for boundary detection, the proposed hybrid network with Cross Entropy loss for boundary detection, the I2I-3D~\cite{merkow2016i2i3d} network, and the state-of-the-art BESNet~\cite{odahirohisa2018besnet}. Table~\ref{tab: results_comp_card} summarized the experimental results and showed that our method outperforms the others in HD95 and ABD. For DSC, our method performs better than all other methods except for BESNet (with a tiny gap).

Fig.~\ref{fig: results_comp} shows a qualitative comparison of the predicted boundaries between our method and the aforementioned methods. Here we choose slices containing a calcified plaque (the upper row) and a non-calcified plaque (the lower row) respectively as representative cases for illustration. For given target slices, the gap between the detection result of morphological Snake and the GT is large compared with other methods. Even though segmentation-based approaches (i.e., Res-UNet with CE and Dice loss, respectively) can obtain a closed outline, they failed to obtain smooth borders. When replacing the proposed loss with a regular Cross Entropy loss, our architecture failed to obtain closed boundaries, and the detected boundaries are too thick. On the other hand, the hybrid network with contour-regularized WHD loss obtains closed boundary in all cases. This verifies the efficiency of the contour-regularized WHD loss on detecting connected boundaries with one-pixel thickness. Our proposal outperformed the Res-UNet with contour-regularized WHD loss in all three metrics. This validates the superiority of the hybrid architecture over a fully-2D approach without considering slice continuity. Our method outperformed the state-of-the-art BESNet and the I2I-3D network in HD95 and ABD with obviously lower score (especially in ABD) and is comparable to BESNet in DSC. In summary, our method comprehensively performed best among the seven methods with both boundary points connectivity and segmentation area consistency.

\begin{table}
\centering
\vspace{-5mm}
\caption{Quantitative results comparisons of our method in terms of DSC, ABD and HD95.} \label{tab: results_comp_card}
\begin{tabular}{l|c|c|c}
\hline
Method &  HD95 & ABD & DSC \\
\hline
morphological Snake~\cite{marquez2014amorph} &  24.79 & 19.44 & 0.56 \\
Res-UNet~\cite{Zhang2018RoadEB} (Seg) +  Dice & 4.59 & 2.37 & 0.82\\
Res-UNet~\cite{Zhang2018RoadEB} (Seg) + CE & 3.43 & 1.62 & 0.85 \\
Res-UNet~\cite{Zhang2018RoadEB} (Bound) + WHD & 3.56 & 1.68 & 0.81 \\
BESNet~\cite{odahirohisa2018besnet} & 3.39 & 2.61  & \textbf{0.86}  \\
I2I-3D~\cite{merkow2016i2i3d} & 4.98 & 2.76 & - \\
hybrid (Bound) + CE  & 4.76 & 3.16  & -  \\ 
hybrid (Bound) + WHD (our) & \textbf{3.20} & \textbf{1.53} & 0.85 \\ 
\hline
\end{tabular}
\vspace{-5mm}
\end{table}

\section{Conclusion} 
We proposed a novel method for coronary artery wall detection in CCTA-CPR scans. The partial volume effect in these scans causes blurry boundaries that hinder the continuity and connectivity of the output boundaries. To solve this, we employed a hybrid network architecture and proposed a novel contour-regularized WHD loss. This allows modeling the relationship between adjacent slices, and provides closed inner and outer boundaries with one-pixel thickness. Experimental results showed that our method outperforms the state-of-the-art and other related works.


\bibliographystyle{IEEEbib}
\bibliography{bibliomed.bib}

\begin{thebibliography}{10}

\bibitem{marquez2014amorph}
P.~{Márquez-Neila}, L.~{Baumela}, and L.~{Alvarez},
\newblock ``A morphological approach to curvature-based evolution of curves and
  surfaces,''
\newblock {\em IEEE Transactions on Pattern Analysis and Machine Intelligence},
  vol. 36, no. 1, pp. 2--17, 2014.

\bibitem{tayel2017449}
Mazhar~B. Tayel, M.A. Massoud, and Y.~Farouk,
\newblock ``A modified segmentation method for determination of iv vessel
  boundaries,''
\newblock {\em Alexandria Engineering Journal}, vol. 56, no. 4, pp. 449 -- 457,
  2017.

\bibitem{tayel2014Anas}
Mazhar~B. Tayel, Magdy~A. Massoud, and Yasser. Shehata,
\newblock ``An automatic segmentation for determination of iv vessel
  boundaries,''
\newblock {\em International Journal of Bioscience, Biochemistry and
  Bioinformatics}, vol. 4, no. 4, pp. 218--223, 2014.

\bibitem{merkow2015strctual}
Jameson Merkow, Zhuowen Tu, David Kriegman, and Alison Marsden,
\newblock ``Structural edge detection for cardiovascular modeling,''
\newblock in {\em Medical Image Computing and Computer-Assisted Intervention -
  MICCAI 2015}, 2015, pp. 735--742.

\bibitem{ronneberger2015unet}
Olaf Ronneberger, Philipp Fischer, and Thomas Brox,
\newblock ``U-{N}et: {C}onvolutional networks for biomedical image
  segmentation,''
\newblock in {\em Medical Image Computing and Computer-Assisted Intervention -
  MICCAI 2015}, 2015, pp. 234--241.

\bibitem{Zhang2018RoadEB}
Zhengxin Zhang, Qingjie Liu, and Yunhong Wang,
\newblock ``Road extraction by deep residual {U}-{N}et,''
\newblock {\em IEEE Geoscience and Remote Sensing Letters}, vol. 15, pp.
  749--753, 2018.

\bibitem{oktay2018attentionunet}
Ozan Oktay, Jo~Schlemper, Lo{\"i}c~Le Folgoc, Matthew C.~H. Lee, Mattias~P.
  Heinrich, Kazunari Misawa, Kensaku Mori, Steven~G. McDonagh, Nils~Y.
  Hammerla, Bernhard Kainz, Ben Glocker, and Daniel Rueckert,
\newblock ``Attention u-net: Learning where to look for the pancreas,''
\newblock {\em ArXiv}, vol. abs/1804.03999, 2018.

\bibitem{Alom2018RecurrentRC}
Md.~Zahangir Alom, Mahmudul Hasan, Chris Yakopcic, Tarek~M. Taha, and
  Vijayan~K. Asari,
\newblock ``Recurrent residual convolutional neural network based on u-net
  (r2u-net) for medical image segmentation,''
\newblock {\em ArXiv}, vol. abs/1802.06955, 2018.

\bibitem{kohl2018probunet}
Simon Kohl, Bernardino Romera-Paredes, Clemens Meyer, Jeffrey De~Fauw,
  Joseph~R. Ledsam, Klaus Maier-Hein, S.~M.~Ali Eslami, Danilo Jimenez~Rezende,
  and Olaf Ronneberger,
\newblock ``A probabilistic u-net for segmentation of ambiguous images,''
\newblock in {\em Advances in Neural Information Processing Systems 31}, pp.
  6965--6975. 2018.

\bibitem{odahirohisa2018besnet}
Hirohisa Oda, Holger~R. Roth, Kosuke Chiba, Jure Sokoli{\'{c}}, Takayuki
  Kitasaka, Masahiro Oda, Akinari Hinoki, Hiroo Uchida, Julia~A. Schnabel, and
  Kensaku Mori,
\newblock ``{BESN}et: {B}oundary-enhanced segmentation of cells in
  histopathological images,''
\newblock in {\em Medical Image Computing and Computer Assisted Intervention -
  MICCAI 2018}, 2018, pp. 228--236.

\bibitem{bian2017convednet}
Cheng Bian, Ran Lee, Yi-Hong Chou, and Jie-Zhi Cheng,
\newblock ``Boundary regularized convolutional neural network for layer parsing
  of breast anatomy in automated whole breast ultrasound,''
\newblock in {\em Medical Image Computing and Computer Assisted Intervention -
  MICCAI 2017}, 2017, pp. 259--266.

\bibitem{merkow2016i2i3d}
Jameson Merkow, Alison Marsden, David Kriegman, and Zhuowen Tu,
\newblock ``Dense volume-to-volume vascular boundary detection,''
\newblock in {\em Medical Image Computing and Computer-Assisted Intervention -
  MICCAI 2016}, 2016, pp. 371--379.

\bibitem{wdhfjavier2018}
Javier Ribera, David G{\"{u}}era, Yuhao Chen, and Edward~J. Delp,
\newblock ``Weighted {H}ausdorff distance: {A} loss function for object
  localization,''
\newblock {\em ArXiv}, vol. abs/1806.07564, 2018.

\bibitem{suzuki85findcontour}
Satoshi Suzuki and Keiichi Abe,
\newblock ``Topological structural analysis of digitized binary images by
  border following.,''
\newblock {\em Computer Vision, Graphics, and Image Processing}, vol. 30, no.
  1, pp. 32--46, 1985.

\end{thebibliography}

\end{document}